\documentclass[12pt,a4paper]{article}
\usepackage{amsmath,amsthm,amsfonts,amssymb,bbm}
\usepackage{graphicx,psfrag,subfigure}
\usepackage{cite}
\usepackage{hyperref}

\newcommand{\E}{\mathbbm{E}}

\newcommand{\R}{\mathbb{R}}

\newcommand{\Z}{\mathbb{Z}}
\newcommand{\dx}{\mathrm{d}}

\newcommand{\Var}{\mathrm{Var}}
\renewcommand{\H}{\mathbb{H}}
\newcommand{\C}{\mathbb{C}}

\DeclareMathOperator*{\GFF}{GFF}

\renewcommand{\Im}{\mathrm{Im}}

\newtheorem{prop}{Proposition}
\newtheorem{thm}[prop]{Theorem}

\title{Anisotropic KPZ growth in $2+1$ dimensions:\\ fluctuations and covariance structure}
\author{Alexei Borodin\thanks{California Institute of Technology and Institute for Information Transmission Problems, Moscow, e-mail: \texttt{borodin@caltech.edu}},
Patrik L. Ferrari\thanks{Institute for Applied Mathematics, Bonn University, \newline
e-mail: \texttt{ferrari@wiener.iam.uni-bonn.de}}}

\date{November 4, 2008}

\begin{document}
\maketitle \sloppy

\begin{abstract}
In~\cite{BF08} we studied an interacting particle system which can
be also interpreted as a stochastic growth model. This model belongs
to the anisotropic KPZ class in $2+1$ dimensions. In this paper we
present the results that are relevant from the perspective of
stochastic growth models, in particular: (a) the surface
fluctuations are asymptotically Gaussian on a $\sqrt{\ln{t}}$ scale
and (b) the correlation structure of the surface is asymptotically
given by the massless field.
\end{abstract}

\section{Introduction}
We consider growth models of an interface described, at least on a
coarse-grained level, by a height function $x\mapsto h(x,t)$, where
$t\in\R$ is the time and $x\in\R^d$ (or $\Z^d$) is the position on a
$d$-dimensional substrate. When the growth is \emph{local}, then on
a macroscopic level the speed of growth $v$ is a function of the
slope only
\begin{equation}
v=v(\nabla h).
\end{equation}
The class of models we consider are \emph{stochastic} with a \emph{smoothing mechanism} leading to a non-random limit shape,
\begin{equation}
h_{\rm ma}(\xi):=\lim_{t\to\infty} \frac{h(\xi t,t)}{t},
\end{equation}
and in the Kardar-Parisi-Zhang (KPZ) class, related to the KPZ
equation
\begin{equation}\label{eq3}
\frac{\partial h}{\partial t}=\nu\Delta h+\frac{\lambda}{2}(\nabla h)^2+\eta,
\end{equation}
with $\eta$ a local space-time noise. (\ref{eq3}) is a mesoscopic equation, in which the smoothing mechanism is modeled by the $\nu\Delta h$ term, which is physically related with the surface tension. $\eta$ is the stochastic part, while the non-linear term $\frac{\lambda}{2}(\nabla h)^2$ is coming from Taylor series of $v(\nabla h)$ around $\nabla h=0$. Indeed, (\ref{eq3}) assumes that one sets the system of coordinates such that $\nabla h$ is small. It is then possible to make the choice so that the 0th and 1st Taylor coefficients vanish. In~\cite{KPZ86} it was argued that for the long time behavior, higher order terms become irrelevant.

Before describing the two-dimensional case, let us spend one
paragraph on the $d=1$ situation. By non-rigorous arguments, the
scaling exponent in $d=1$ were correctly derived to be $1/3$ for the
fluctuations and $2/3$ for the spatial correlations,
see~\cite{FNS77,BKS85,BS95}. This means that the height function $h$
rescaled as
\begin{equation}\label{eq4}
h_t^{\rm resc}(u):=\frac{h(\xi t+u t^{2/3},t)-t h_{\rm ma}\left((\xi t+u t^{2/3})/t\right)}{t^{1/3}},
\end{equation}
in the long time limit will converge to a well-defined stochastic
process. The analysis of simplified models in the KPZ class gives
the following results for noise-free initial conditions (results
expected to hold for the whole class by universality). If $h_{\rm
ma}$ is straight (i.e.~second derivative is zero) around $\xi$, then
the limit process is the Airy$_1$ process~\cite{BFP06,BFPS06}, if
$h_{\rm ma}$ is curved, the process is the Airy$_2$
process~\cite{Sas05,PS02b,Jo00b,Jo03b}. Further results as the
transition between these processes~\cite{BFS07} or correlations at
different times~\cite{SI07,BF07,BFS07b} with also some unexpected
results~\cite{Fer08} are also available.
See~\cite{Fer07,FP05,De06,PS00b} for reviews
and~\cite{Jo03,BS06,BJ08,QV06,NS04,PS01,FS05a} for related works.

The main reason of this paper is to present results on the
anisotropic class in $d=2$. Going back to the KPZ equation, one
realizes that in two dimensions one should write
\begin{equation}\label{eq5}
\frac{\partial h}{\partial t}=\nu\Delta h+ Q(\nabla h)+\eta,
\end{equation}
where $Q$ is a quadratic form. It turns out that there are two
classes, depending on the signature of the quadratic form $Q$.

\medskip

\textbf{(a) Isotropic KPZ:} ${\rm sign}(Q)=(+,+)$ or $(-,-)$, i.e.,
the Hessian of $v=v(\nabla h)$ has two eigenvalues of the same sign.
In this case, the fluctuations grow as $t^\beta$ for some $\beta>0$,
but renormalization arguments used in the $d=1$ case fail to give an
answer. Numerical studies on discrete models belonging to the KPZ
class~\cite{FT90,TFW92} as well as numerical solution of the
equation itself~\cite{MKW91}, indicate that $\beta=0.240\pm 0.001$
(ruling out the value $1/4$). For this case, essentially no analytic
results are available.

\medskip

\begin{figure}[t!]
\begin{center}
\includegraphics[angle=90,height=6cm]{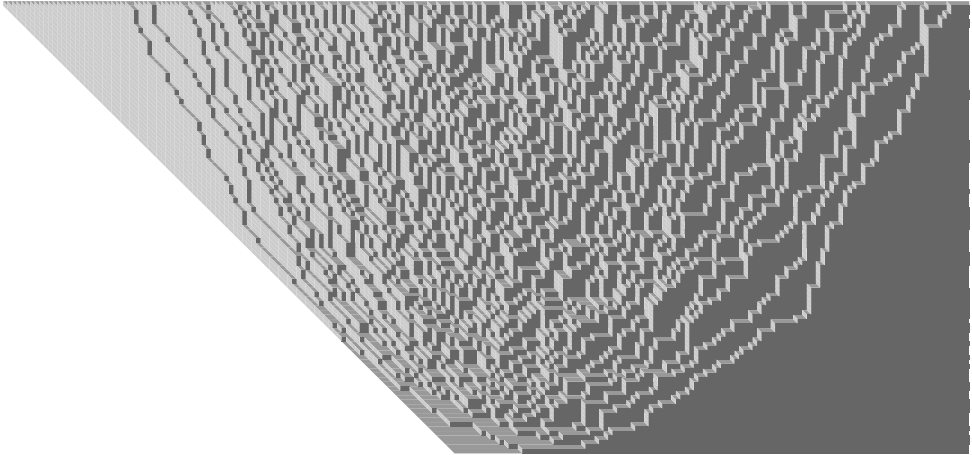}
\caption{A configuration of the model analyzed with $N=100$ particles at time $t=25$. In~\cite{FerAKPZ} there is a Java animation of the model.}
\label{FigSimulazione}
\end{center}
\end{figure}

\textbf{(b) Anisotropic KPZ (=AKPZ):} ${\rm sign}(Q)=(+,-)$. In this
situation, the behavior is quite different and the fluctuations
increase much slower, only as $\sqrt{\ln{t}}$. The first time the
AKPZ was considered was by Wolf in~\cite{Wol91}, who was interested
in growth on vicinal surfaces. They are surfaces with a small tilt
with respect to a high-symmetry plane of the crystalline structure
of the solid, and they are intrinsically asymmetric. By detailed
one-loop computations, Wolf deduced that the non-linearity should be
irrelevant for the roughness, which should grow in time as $\ln(t)$,
i.e.,
\begin{equation}\label{eq6}
\Var(h(\xi t,t))\sim \ln(t),\quad \textrm{ as }t\to\infty,
\end{equation}
exactly as in the $\lambda=0$ case (Edwards-Wilkinson
model~\cite{EW82}). On the numerical side, this drastic difference
have been tested positively soon after~\cite{HHA92} on the model
used by Wolf directly, but also on a deposition model~\cite{KKK98}.
On the analytic side, Pr\"ahofer and Spohn~\cite{PS97} considered a
microscopic model in the AKPZ class and reproduced Wolf's
prediction, namely (\ref{eq6}). Moreover, they computed the local
correlations obtaining
\begin{equation}\label{eq7}
\lim_{t\to\infty}\Var(h(x,t)-h(x',t))\sim \ln|x-x'|,
\end{equation}
for large $|x-x'|\to\infty$, but not growing with $t$.

In our recent work~\cite{BF08} we consider a growth model in the
AKPZ class, for which we determine the detailed correlation
structure on a macroscopic scale. Below we explain the model and
present the results relevant under the perspective of growth models
(see Figure~\ref{FigSimulazione} for an illustration of the surface
we analyzed).

\section{The model}

\begin{figure}[t!]
\begin{center}
\psfrag{x}[b]{$x_1$}
\psfrag{y}[lb]{$x_2$}
\psfrag{z}[b]{$x_3$}
\includegraphics[height=4cm]{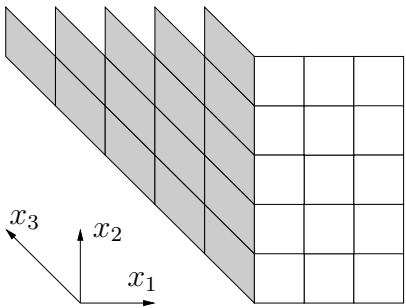}
\caption{Initial condition. The corner of the crystal is at position $(0,0,0)$.}
\label{FigInitialCondition}
\end{center}
\end{figure}

We consider a simplified model of anisotropic evaporation of a
crystal. As initial condition we have a perfect corner on the
positive octant, see Figure~\ref{FigInitialCondition}.
The dynamics is as follows. The evaporation occurs by detaching of
columns along the $x_2$-direction. Each cube at $(x_1,x_2,x_3)$ such
that both its $x_3$-facet and $x_1$-facet are visible, evaporates
with rate one, and by doing so it takes away also all the cubes in
the $x_2$-column above it: $\{(x_1,x_2',x_3)\textrm{ s.t. }
x_2'>x_2\}$. This dynamics is intrinsically asymmetric, since the
$x_2$ and $x_3$ directions play different roles.

\begin{figure}[t!]
\begin{center}
\includegraphics[height=8cm]{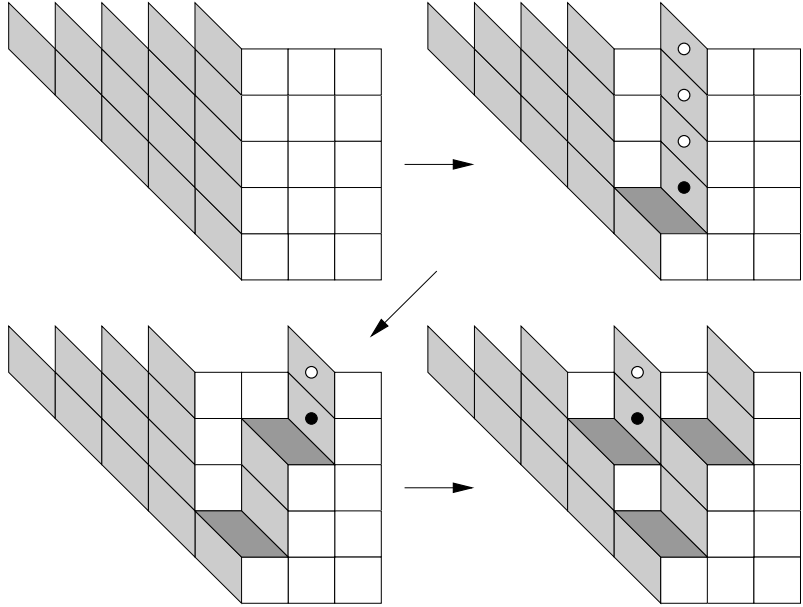}
\caption{An example of the first three moves. The moved cubes are
the ones with the black circles, and also the ones with white
circles which are moved because they are in a column above the black
ones.} \label{FigDynamics}
\end{center}
\end{figure}

There are several possible choices of coordinate which can be used to describe the interface as in Figure~\ref{FigSimulazione}. The one we have chosen in~\cite{BF08} is the following. Consider the two-dimensional projection as in Figure~\ref{FigSimulazione}, whose points can be described by $(x_1-y,y=x_3+x_2)$. For convenience, we take $x=x_1$ as the first coordinate and $y$ as the second one, while the height function will be given by $x_3$:
\begin{equation}
(x,y,h_t(x,y))=(x_1,x_2+x_3,x_3).
\end{equation}
The surface grows with a non-zero speed. Thus we rescale our coordinates as $x_1=\xi t$, $x_2=\zeta t$, $x_3=h_t(\xi t,\zeta t)$ so that $x=\xi t$, $y=\zeta t-h_t(\xi t,\zeta t)$. Moreover, in the long time limit, our surface has a \emph{deterministic limit shape}
\begin{equation}\label{eqMacroH}
\textit{\textbf{h}}(\xi,\zeta):=\lim_{t\to\infty} \frac{h_t(\xi t,\zeta t)}{t}.
\end{equation}
The limit shape consists of facets joined by a curved region in the coordinate range
\begin{equation}\label{eqBulk}
(1-\sqrt{\zeta})^2 < \xi < (1+\sqrt{\zeta})^2.
\end{equation}
Explicitly, for $(\xi,\zeta)$ in the curved region,
\begin{equation}\label{eqh}
\textit{\textbf{h}}(\xi,\zeta)=\frac{1}{\pi}\int_{\xi}^{(1+\sqrt{\zeta})^2} \arccos\left(\frac{1+\zeta-\xi'}{2\sqrt{\zeta}}\right)\dx \xi'.
\end{equation}
In the curved region, in the long time limit, one locally sees a flat interface characterized by the two slopes
\begin{equation}\label{eq8}
\textit{\textbf{h}}_\xi :=\frac{\partial \textit{\textbf{h}}}{\partial \xi},\quad \textit{\textbf{h}}_\zeta :=\frac{\partial \textit{\textbf{h}}}{\partial \zeta}.
\end{equation}
The two slopes determine uniquely a stationary measure, and since our \emph{growth is short range}, the speed of growth $v$ will be a function of $\textit{\textbf{h}}_\xi $ and $\textit{\textbf{h}}_\zeta $ only. Indeed, we get
\begin{equation}
v=v(\textit{\textbf{h}}_\xi ,\textit{\textbf{h}}_\zeta )=-\frac{1}{\pi}\frac{\sin(\pi \textit{\textbf{h}}_\xi )\sin(\pi \textit{\textbf{h}}_\zeta )}{\sin(\pi (\textit{\textbf{h}}_\xi +\textit{\textbf{h}}_\zeta ))}.
\end{equation}
From this expression, we can compute the determinant of the Hessian of $v$ with the result
\begin{equation}
\left|\begin{array}{cc}
\partial_{\textit{\textbf{h}}_\xi }\partial_{\textit{\textbf{h}}_\xi }v & \partial_{\textit{\textbf{h}}_\xi }\partial_{\textit{\textbf{h}}_\zeta }v \\ \partial_{\textit{\textbf{h}}_\zeta }\partial_{\textit{\textbf{h}}_\xi }v & \partial_{\textit{\textbf{h}}_\zeta }\partial_{\textit{\textbf{h}}_\zeta }v
\end{array}\right| = -4\pi^2\frac{\sin(\pi \textit{\textbf{h}}_\xi )^2\sin(\pi \textit{\textbf{h}}_\zeta )^2}{\sin(\pi(\textit{\textbf{h}}_\xi +\textit{\textbf{h}}_\zeta ))^4} <0
\end{equation}
for $\textit{\textbf{h}}_\xi ,\textit{\textbf{h}}_\zeta
,\textit{\textbf{h}}_\xi +\textit{\textbf{h}}_\zeta \in (0,1)$
(which is equivalent to being in the curved region).

Thus, our stochastic growth model has a smoothing mechanism
determining deterministic limit shape, the growth rule is local, and
the Hessian has eigenvalues of opposite signs. Therefore, our model
belongs to the anisotropic KPZ class (AKPZ) in $2+1$ dimensions.

\section{Results}
\subsubsection*{Logarithmic fluctuations}
The first result concerns Wolf's prediction of the logarithmic
growth of the fluctuations. In Theorem 1.2 of~\cite{BF08} we prove
the following result.
\begin{thm}\label{thm1}
Consider any macroscopic point in the curved region, i.e.,
$(x,y)=(\xi t,\zeta t)$ with $(\xi,\zeta)$ satisfying
(\ref{eqBulk}). Then, the random variable
\begin{equation}\label{eq14}
\widetilde h_t(\xi,\zeta):=\frac{h_t(x,y)-\E(h_t(x,y))}{\sqrt{\ln{t}}}
\end{equation}
converges in the $t\to\infty$ limit to a normal random variable with mean zero and variance $1/(2\pi^2)$.
\end{thm}

\subsubsection*{Complex structure and local measure}
\begin{figure}[t!]
\begin{center}
\psfrag{Omega}[c]{$\Omega$}
\psfrag{0}[c]{$0$}
\psfrag{1}[c]{$1$}
\psfrag{zeta}[r]{$\sqrt{\zeta}$}
\psfrag{xi}[l]{$\sqrt{\xi}$}
\psfrag{pixi}[c]{$\pi_\xi$}
\psfrag{pizeta}[c]{$\pi_\zeta$}
\psfrag{pi1}[c]{$\pi_1$}
\includegraphics[height=4cm]{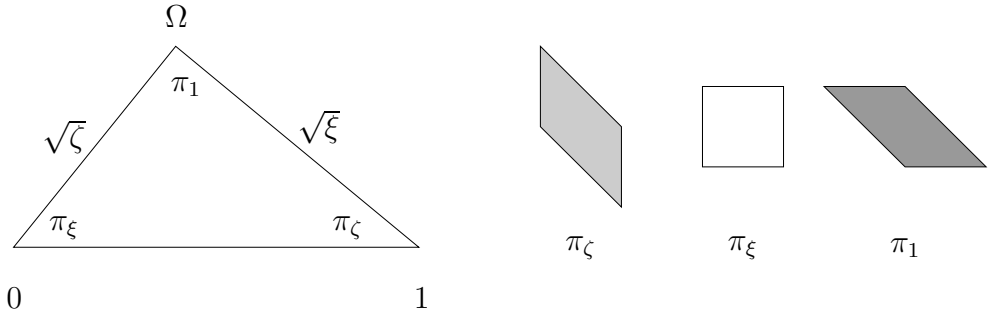}
\caption{Map $\Omega$ and types of facets with associated angles.}
\label{FigGeometry}
\end{center}
\end{figure}

To present the other results, we need to define the following
mapping. For any $(\xi,\zeta)$ satisfying (\ref{eqBulk}), define the
map $\Omega$ to the upper (complex) half-plane $\H:=\{z\in\C\textrm{
s.t. }\Im(z)>0\}$ as follows:
\begin{equation}
|\Omega|=\sqrt{\zeta},\quad |1-\Omega|=\sqrt{\xi}.
\end{equation}
Geometrically, the point $\Omega$ is the intersection in $\H$ of the circles\footnote{The notation $B(x,r)$ denotes the circle centered at $x$ of radius $r$.} $B(0,\sqrt{\zeta})$ and $B(1,\sqrt{\xi})$ (see Figure~\ref{FigGeometry}).
Also, denote the angles of the triangle with vertices $0,1,\Omega$ by
\begin{equation}
\pi_\zeta=-\arg(1-\Omega),\quad \pi_\xi=\arg(\Omega),\quad \pi_1=\pi-\pi_\zeta-\pi_\xi.
\end{equation}

Now we discuss the local measure in the curved part of the surface.
If one focuses around macroscopic point $(\xi,\zeta)$,
asymptotically one sees a translation invariant measure
parameterized by the gradient of the surface. The surface can be
also viewed as a random tiling of the plane by three types of
lozenges indicated in Figure~\ref{FigGeometry}. Then, the measure
one sees locally is the unique translation invariant Gibbs measure
on lozenge tilings with prescribed proportions of lozenges
$(\pi_\zeta/\pi,\pi_\xi/\pi,\pi_1/\pi)$
(see~\cite{KOS03,KenLectures,BS08} and references therein for
discussions of these Gibbs measures). From the tiling point of view,
it is easy to understand the connection between the angles
$\pi_\xi$, $\pi_\zeta$ and the asymptotic slopes
$\textit{\textbf{h}}_\xi ,\textit{\textbf{h}}_\zeta$:
\begin{equation}\label{eq18}
\textit{\textbf{h}}_\xi =-\pi_\zeta/\pi,\quad \textit{\textbf{h}}_\zeta =1-\pi_\xi/\pi.
\end{equation}
Indeed, along the $\xi$-direction, the height changes whenever there
are lozenges associated to $\pi_\zeta$, while along the $\zeta$
direction it changes whenever there are lozenges not associated with
$\pi_\xi$ (varying $\zeta$ with fixed $\xi$ corresponds to moving
along the $(1,-1)$ direction in our figures). Finally, the growth
velocity takes a very simple form in terms of $\Omega$:
\begin{equation}\label{eqSpeed}
v=\Im(\Omega)/\pi.
\end{equation}

Interestingly enough, although we study a stochastic growth model
(out of equilibrium), the limit shape is related to a variational
problem appearing for uniformly distributed lozenge tilings of a
given domain. The map $\Omega$ satisfies the complex Burgers
equation
\begin{equation}\label{eqCBE}
(1-\Omega)\frac{\partial \Omega}{\partial \xi}=\Omega \frac{\partial \Omega}{\partial \zeta},
\end{equation}
which has been identified as the Euler-Lagrange equation for the surface free energy for fixed domains, see~\cite{KO07} and references therein. The limit shape gives a solution of (\ref{eqCBE}) and it is encoded in the equation
\begin{equation}\label{eqLS}
\xi\Omega+\zeta(1-\Omega)=\Omega(1-\Omega),\quad \Omega\in\H,\,\xi,\zeta\geq 0.
\end{equation}
Indeed, for any $\Omega\in\H$, there is a unique couple $(\xi,\zeta)\in\R_+^2$ satisfying (\ref{eqLS}), and then the limit shape $\textit{\textbf{h}}(\xi,\zeta)$ can be simply expressed as the integral of the slope as in (\ref{eqh}).

\subsubsection*{Macroscopic correlations}
Finally, let us discuss the covariance structure of different
macroscopic points in the curved part of the surface. At first
sight, in view of (\ref{eq14}) one might think that we have to
compute $\E(\prod_{i=1}^m \widetilde h_t(\xi_i,\zeta_i))$ for
distinct $(\xi_i,\zeta_i)$ satisfying (\ref{eqBulk}). However, this
turns out to converge to zero in the $t\to\infty$ limit.

Our result is that, after mapping by $\Omega$, the height function
converges to the massless field on $\H$, which we denote $\GFF$. The
value of $\GFF$ at a point cannot be defined. However, one can think
of expectations of products of values of $\GFF$ at different points
as being finite and equal to
\begin{multline}\label{eqGFFmoments}
\E[\GFF(z_1)\cdots\GFF(z_m)]\\
=\left\{
\begin{array}{ll}
0&\textrm{ if }m\textrm{ is odd}, \\
\sum\limits_{\textrm{pairings }\sigma} {\cal G}(z_{\sigma(1)},z_{\sigma(2)})\cdots{\cal G}(z_{\sigma(m-1)},z_{\sigma(m)}) &\textrm{ if }m\textrm{ is even},
\end{array}
\right.
\end{multline}
where
\begin{equation}
\E[\GFF(z)\GFF(w)]={\cal G}(z,w):=-\frac{1}{2\pi}\ln\left|\frac{z-w}{z-\bar w}\right|,
\end{equation}
is the Green function of the Laplace operator on $\H$ with Dirichlet boundary conditions. In (\ref{eqGFFmoments}) one recognizes Wick formula for Gaussian processes. For further details on the massless field (also called Gaussian Free Field), see e.g.~\cite{She03}.

Our result on the correlations on a \emph{macroscopic} scale is as
follows (Theorem 1.3 in~\cite{BF08}).
\begin{thm}\label{thm2}
For any $m=1,2,\ldots$, consider any $m$ distinct pairs
$\{(\xi_i,\zeta_i), i=1,\ldots,m\}$ which satisfy (\ref{eqBulk}).
Denote
\begin{equation}
H_t(\xi,\zeta):=\sqrt{\pi}\left[h_t(\xi t,\zeta t)-\E(h_t(\xi t,\zeta t))\right]
\end{equation}
and $\Omega_i:=\Omega(\xi_i,\zeta_i)$. Then,
\begin{equation}\label{eqCorr}
\lim_{t\to\infty} \E(H_t(\xi_1,\zeta_1)\cdots H_t(\xi_m,\zeta_m)) = \E[\GFF(\Omega_1)\cdots \GFF(\Omega_m)]
\end{equation}
\end{thm}

At first sight it might look strange that in Theorem~\ref{thm1} we
have $\sqrt{\ln{t}}$ scaling, while in Theorem~\ref{thm2} it is not
present. If we consider $H_t$ as random variable, Theorem~\ref{thm1}
tells us that, in the $t\to\infty$ limit, $H_t$ does not converge to
a well-defined random variable (the $\sqrt{\ln{t}}$ normalization is
missing). On the other hand, Theorem~\ref{thm2} tells us that the
macroscopic correlations of $H_t$ are non-trivial. The reason is
that asymptotically the interface $H_t$ will take values in the
space of distributions, not ordinary functions.

Our Theorem 1.3 in~\cite{BF08} is actually more general. Indeed, we
obtain the limiting correlations also for the height function at
different times, provided that the space-time events at which we
consider the height function satisfy a ``space-like condition''.
When we consider heights at different times, we cannot use $t$
anymore as a large parameter. We use $L$ instead, set $t=\tau L$,
and denote by $(\nu L,\eta L,\tau L)$ a space-time point ($\nu=\tau
\xi$, $\eta=\tau \zeta$). Then, the condition (\ref{eqBulk}) that a
point is in the curved region becomes
\begin{equation}\label{eqBulkExt}
(\sqrt{\tau}-\sqrt{\eta})^2<\nu<(\sqrt{\tau}+\sqrt{\eta})^2.
\end{equation}
Then, the mapping $\Omega$ is now from a triple $(\nu,\eta,\tau)$ to $\H$ given by the condition
\begin{equation}
|\Omega|=\sqrt{\eta/\tau},\quad |1-\Omega|=\sqrt{\nu/\tau}.
\end{equation}
Then the full result of our Theorem 1.3 in~\cite{BF08} is the following.
\begin{thm}\label{thm3}
For any $m=1,2,\ldots$, consider any $m$ distinct triples $\{(\nu_i,\eta_i,\tau_i), i=1,\ldots,m\}$, which satisfy (\ref{eqBulkExt}) and the space-like condition
\begin{equation}\label{eqSpaceLike}
\tau_1\leq \tau_2\leq \ldots\leq \tau_m,\quad \eta_1\geq \eta_2\geq\ldots\geq \eta_m.
\end{equation}
Denote
\begin{equation}
H_L(\nu,\eta,\tau):=\sqrt{\pi}\left[h_{\tau L}(\nu L,\eta L)-\E(h_{\tau L}(\nu L,\eta L))\right]
\end{equation}
and $\Omega_i:=\Omega(\nu_i,\eta_i,\tau_i)$. Then,
\begin{equation}\label{eqCorr2}
\lim_{L\to\infty} \E(H_L(\nu_1,\eta_1,\tau_1)\cdots H_L(\nu_m,\eta_m,\tau_m)) = \E[\GFF(\Omega_1)\cdots\GFF(\Omega_m)].
\end{equation}
\end{thm}

There is an interesting fact to notice. Consider for simplicity the following two pairs of space-time points
\begin{equation}
\{(\nu_1,\eta_1,1),(\nu_2,\eta_2,1)\textrm{ with }\eta_2<\eta_1\}
\end{equation}
and
\begin{equation}
\{(\nu_1,\eta_1,1),(\nu',\eta',\tau) \textrm{ s.t. }\Omega(\nu_2,\eta_2,1)=\Omega(\nu',\eta',\tau), \tau>1\}.
\end{equation}
Then, Theorem~\ref{thm3} tell us that the correlations of $H_L$ at these two pairs are asymptotically the same. Also, remark that space-time points have the same image by $\Omega$ on $\H$ if and only if they have the same macroscopic slopes defined in (\ref{eq8}), see (\ref{eq18}).

This a priori unexpected phenomenon, can be explained as follows.
Along the space-time curves with constant $\Omega$, the height
decorrelates on a much slower time scale than $L$. The same fact has
been already observed in KPZ growth on a $d=1$
substrate~\cite{Fer08} (there the correct scale is $L$ instead of
$L^{2/3}$ for spatial correlations).

As a consequence of this interpretation, we conjecture that
Theorem~\ref{thm3} holds \emph{without the space-like condition}
(\ref{eqSpaceLike}), provided that $\Omega_1,\ldots,\Omega_m$ remain
pairwise distinct.


\end{document}